# A novel model and estimation method for the individual random component of earthquake ground-motion relations

Mathias Raschke, Freelancer, Stolze-Schrey Str.1, 65195 Wiesbaden, Tel. +49 611 98819561, Email: mathiasraschke@t-online.de

In this paper, I introduce a novel approach to modelling the individual random component (also called the intra-event uncertainty) of a ground-motion relation (GMR), as well as a novel approach to estimating the corresponding parameters. In essence, I contend that the individual random component is reproduced adequately by a simple stochastic mechanism of random impulses acting in the horizontal plane, with random directions. The random number of impulses was Poisson distributed. The parameters of the model were estimated according to a proposal by Raschke (2013a), with the sample of random difference $\xi=ln(Y_1)-ln(Y_2)$, in which $Y_1$ and $Y_2$ are the horizontal components of local ground-motion intensity. Any GMR element was eliminated by subtraction, except the individual random components. In the estimation procedure the distribution of difference $\xi$ was approximated by combining a large Monte Carlo simulated sample and Kernel smoothing. The estimated model satisfactorily fitted the difference $\xi$ of the sample of peak ground accelerations, and the variance of the individual random components was considerably smaller than that of conventional GMRs. In addition, the dependence of variance on the epicentre distance was considered; however, a dependence of variance on the magnitude was not detected. Finally, the influence of the novel model and the corresponding approximations on PSHA was researched. The applied approximations of distribution of the individual random component were satisfactory for the researched example of PSHA.

*ground-motion relation, individual random component, intra-event uncertainty, probabilistic seismic hazard analysis*

## 1 Introduction

The probabilistic seismic hazard analysis (PSHA) estimates the local seismic impact for the defined average of the annual exceedance frequency (equal to the reciprocal of the return period). The ground-motion relation (GMR) is an important element of PSHA that describes the relation between the local ground-motion intensity and the event parameters, such as the magnitude. Another term for ground-motion relation is ground-motion prediction equation; however, in this paper, the term GMR (Atkinson, 2006; Raschke, 2013a) is preferred. The reasons are that GMR includes more elements than one equation and regression analysis does not have to be applied directly. The term 'prediction equation' is related to the regression analysis, which is commonly used for estimating the parameters of GMR (e.g., Strasser et al., 2009), with event parameters and source distances as predictors. Douglas (2011) provided a good overview of GMRs published before 2010. The physical unit of local ground-motion intensity *Y* is the peak ground acceleration (PGA), or the maximum of another type of local time history.

The basic formulation for GMR is

$$Y = \varepsilon_a \varepsilon_b\, g(\mathbf{X}), \quad E(Y) = g(\mathbf{X}), E(\varepsilon_a) = E(\varepsilon_b) = 1 \tag{1a}$$

and

$$\ln(Y) = \ln(\varepsilon_a) + \ln(\varepsilon_b) + \ln(g(\mathbf{X})) \tag{1b}$$

These equations also constitute the basic formulation of statistical regression models (cf. Rawlings et al., 1998; Fahrmeier et al., 2013), with the regression function (prediction equation) *g(**X**)*. It is called 'expectation function' here because it describes an expectation and regression analysis is not applied. The expectation *E(.)* and the variance *V(.)* are the important moments of random variables, and are used for parameterization. The predicting vector of the expectation function is **X** and it includes variables such as the event parameters (e.g., the magnitude), the distance between the source and the site, and the site effects. The individual random component $\varepsilon_a$ is a random realisation for every specific location and single event. The random component $\varepsilon_b$ has only one realisation per event and is the same for every location. The random component $\varepsilon_a$ is also called the intra-event uncertainty or variability, while $\varepsilon_b$ is also called the inter-event uncertainty or variability (e.g., Campbell and Bozorgnia, 2008).

The spreading of the random components is quantified by the variances *V(ε)* and the sigma $\sigma = \sqrt{V(\ln(\varepsilon))}$, which considerably influences the results of PSHA. Since this estimation is most important, various researchers have considered the aspect in detail (e.g., Abrahamson and Silva, 2008; Atkinson, 2004; Bommer et al., 2007; Campbell and Bozorgnia, 2008). In addition, the aspect of random components has been the subject of dedicated research and discussions (e.g., Atkinson, 2006; Bommer and Abrahamson, 2006; Al Atik et al., 2010).

The parameters of Eq. (1b) are estimated for most GMR by regression analysis, with the use of a least squares estimation, assuming a normal distribution for *ln(ε)* that implies a log-normally distributed *ε*. The variance of the random component is equal to the residual variance of the regression analysis. However, arguments have been presented against the assumption of the log-normal distribution for the individual random component $\varepsilon_a$ (Dupuis and Flemming, 2006; Raschke, 2013a; Pavlenko, 2015). Furthermore, the residual variance is not an appropriate estimator for *V(ε)* because of the principal of area equivalence, as was revealed by Raschke (2013a). According to this principle, there is the actual function *g(**X**)* of Eq. (1), as well as that of the estimated function *g*<sup>\*</sup>*(**X**)*. These functions are not identical, which is brought about by various factors, such as the estimation error and the finite sample. However, the functions can be area equivalent, as indicated by the schematic example of an event (Fig. 1a) in a one-dimensional geographical space. The estimation $g^*(X)$ includes local biases, as indicated in Fig. 1b. The over- and underestimated locations include an equilibrium determined by the area equivalence, as illustrated by the areas $A$ of $g(X)$ and $A^*$ of $g^*(X)$ (see Fig. 1a). These areas include all the points to which $g(X) \geq 1.5 \text{m/s}^2$ or $g^*(X) \geq 1.5 \text{m/s}^2$ apply, and are the same size, which also applies to any threshold because $g(X)$ and $g^*(X)$ are area equivalent. Consequently, the GMRs exert equivalent influence on the hazard of all the locations in the geographical space of PSHA, which would result in the equal average functions of the annual exceedance rates of the local ground-motion intensities in a PSHA (cf. Raschke, 2014).

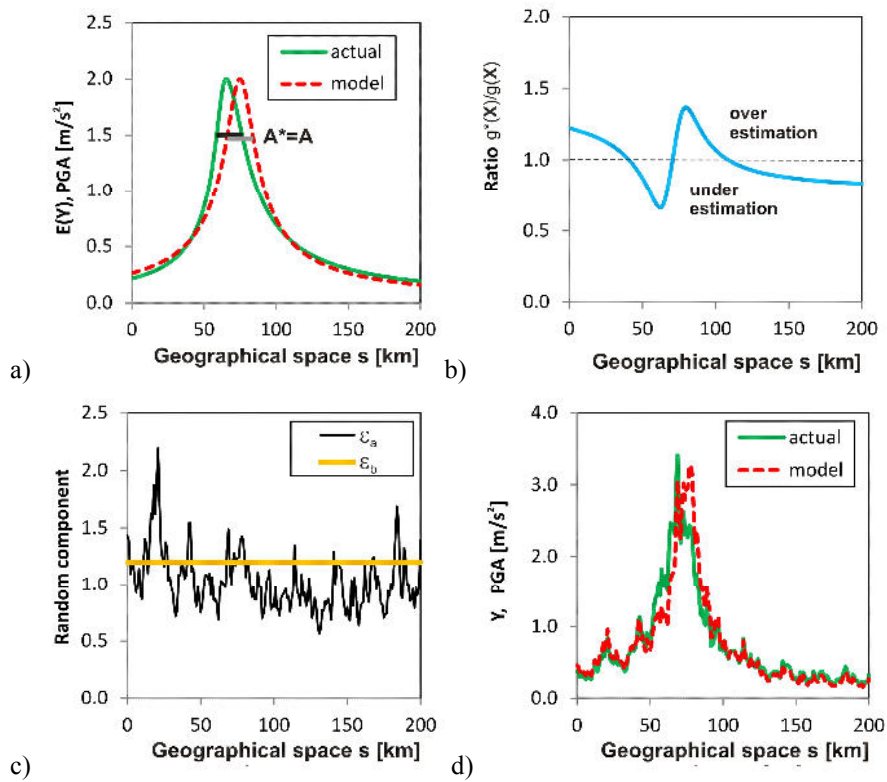

Fig. 1: The principle of area equivalent GMRs explained by a fictitious schematic example in an one-dimensional geographical space, with coordinate $s$: a) Actual expectation function $g(X)$ and modelled expectation function $g^*(X)$, b) Local biases of $g^*(X)$, c) Realisations of the random components ($\varepsilon_a$ with a spatial correlation), d) Realisation of $Y(s)$.

The random components $\varepsilon_a$ and $\varepsilon_b$ are realisations in the geographical space (Fig. 1c) and are simply added to the expectation functions, according to Eq. (1) (Fig. 1d).

The problem is that in the regression analysis for recent GMRs, the differences between $g(X)$ and $g^*(X)$ are interpreted as parts of the random component $\varepsilon_a$. However, this leads to an overestimation of the dispersion measures $V(\varepsilon_a)$ and $\sigma_a = \sqrt{V(\ln(\varepsilon_a))}$ and, consequently, results in a global overestimation of the seismic hazard (cf. Raschke, 2013a). I particularly emphasise that a global overestimation of $V(\varepsilon_a)$ and $\sigma_a$ cannot compensate for the local biases of Fig. 1b. Accordingly, an alternative estimation for $V(\varepsilon_a)$ and $\sigma_a$ was needed and was developed, as described in this paper.

Raschke (2013a) has suggested estimating the parameters of $\varepsilon_a$ by determining the sample of the random difference $\xi$ by:

$$\xi = \ln(Y_1/Y_2) = \ln(\varepsilon_{a,1}\varepsilon_b\, g(\mathbf{X})) - \ln(\varepsilon_{a,2}\varepsilon_b\, g(\mathbf{X})) = \ln(\varepsilon_{a,1}) - \ln(\varepsilon_{a,2}),\ E(\xi) = 0 \qquad (2)$$

where $Y_1$ and $Y_2$ are the two horizontal components of the local ground-motion intensity, and $\varepsilon_{a,1}$ and $\varepsilon_{a,2}$ are the corresponding individual random components. The orientations of the components are perpendicular to each other. The advantage of this approach is that any information or estimation of $g(\mathbf{X})$ and $\varepsilon_b$ of Eq. (1) is not required. In this paper, I expound on this concept and I construct and research a random mechanism that generates the random components $\varepsilon_{a,1}$ and $\varepsilon_{a,2}$. The result is the distribution model of $\xi$ equal to an empirical distribution, and the generating mechanism of $\xi$ is applied to estimate the distribution parameters of $\varepsilon_a$.

The details of the new approach are presented in the following section, while in section 3 a sample of empirical data is introduced. This sample of data was analysed and the estimation of the variance of $\varepsilon_{a,1}$ and $\varepsilon_{a,2}$ is presented in section 4. In section 5, the results of the research on the possible effects on PSHA is provided, including approximations for the distribution models of $\varepsilon_a$. Finally, my conclusions and a discussion of the results are presented, as well as proposed further research.

## 2  The random mechanism generating $\varepsilon_a$

A random mechanism was constructed that approximates the generation of the individual random components of Eqs (1–2). For this purpose, I researched the ground-motion intensity $Y$, which is an absolute maximum of the time history for a fixed orientation (direction) angle $w$. An example of an earthquake time history of ground acceleration is shown as a polar plot in Fig. 2a, in which $Y(w)$ is the red line. It is clear that $Y(w)=Y(w+\pi)$.

A similar figure is determined by two simple autoregressive processes (one for each horizontal component), with autocorrelation $r=0.85$ per time step, as shown in Fig. 2b. The similarity is based on the extreme value statistics, with the peak over threshold (POT) analysis (cf. Coles, 2001; Raschke, 2013b). The important extremes of a time history are separated by a simple procedure, namely, filtering the peak over an (appropriately) defined threshold and filtering the maxima of each cluster of POTs. The clusters are partial sequences of the time history.

The POT approach for one variable can be adapted to the two components of an earthquake time history (the green broken lines in Figs 2a and b). Each cluster maxima acts as a random impulse $Z>0$, which is depicted by Fig. 2c, with a random direction $v$. The resulting random component $\varepsilon_a(w)$ is determined by the maxima of a sequence of such impulses, according to Eq. (3a) and Fig. 2d. The random size of this sequence is $k$, and any $Y(w)$ is determined by Eq. (3b).

$$\varepsilon_a(w) = \max(Z_1|\cos(w-v_1)|, Z_2|\cos(w-v_2)|, ..., Z_i|\cos(w-v_i)|, ..., Z_k|\cos(w-v_k)|) \qquad (3a)$$

and

$$Y(w) = \varepsilon_a(w)\varepsilon_b g(\mathbf{X}) \qquad (3b)$$

The axes of the diagrams in Fig. 1 are the scales of $Y_1 = Y(w=0)$ and $Y_2 = Y(w=\pi/2)$. The random orientation $v$ is uniformly distributed with $0 < v \leq 2\pi$. The Poisson distribution (cf. appendix) is preferred for the random integer number $k$, because a Poisson distribution approximates the number of POTs in extreme value statistics (cf. Falk et al., 2011; Poisson approximation). Furthermore, the Poisson distribution has only one parameter, the intensity $\lambda = E(k) = V(k)$. The instance $k=0$ is ignored here, and only results in an extremely small bias if $\lambda$ is not particularly small.

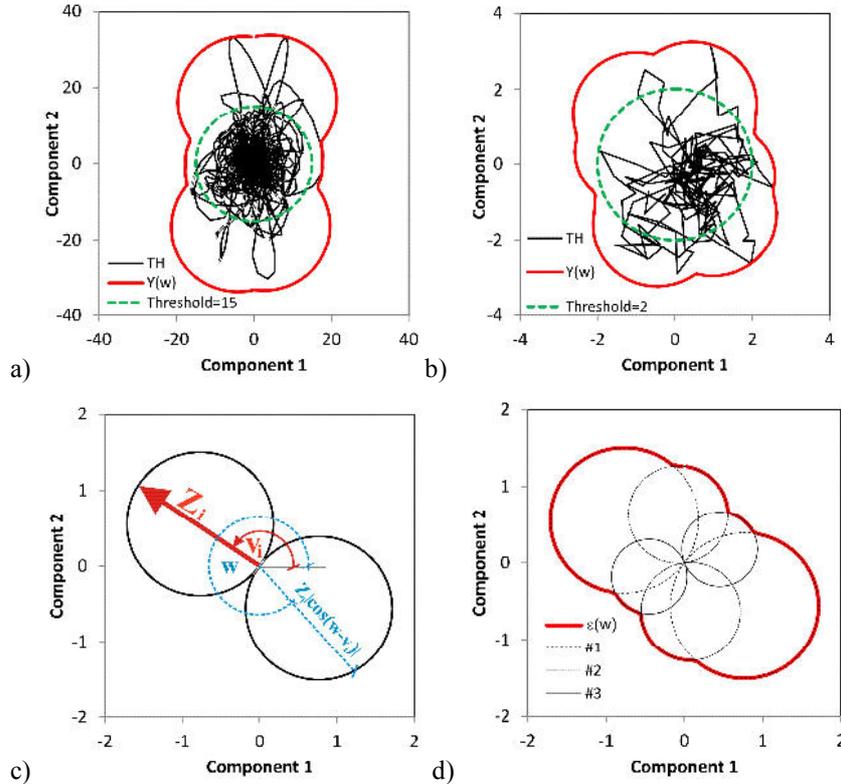

Fig. 2: The concept of the random impulse and the orientation-dependent absolute maximum $Y(w)$: a) Observed time history (TH) of station FKS013 in Japan (NIED, 2015; Record Time 2001/10/02 17:20:12) and corresponding orientation-dependent $Y(w)$, b) Simulated TH of an autocorrelated process, with autocorrelation $r=0.85$ (cf. Upton and Cook, 2008) and the resulting $Y(w)$, c) Random impulse and random direction, d) Example of $\varepsilon_a(w)$ as a result of a sequence of random impulses with $k=3$.

The appropriate distribution of the random variable $Z$ is probably an extreme value distribution, because it represents the maximum of a partial time series/cluster. A Gumbel distribution, with an extreme value index $\gamma=0$, is assumed here, which corresponds with the estimation of Dupuis and Fleming (2006), with $\gamma \approx 0$ for the individual random component of GMR. Alternative distributions for $Z$ should be considered to evaluate the sensitivity of the approach.

The parametrisation of the described stochastic mechanism is as follows. There are fixed expectations $E(\xi)=0$, $E(\varepsilon_{a,1})= E(\varepsilon_{a,2})=1$. The variable parameters are $E(Z)$, $V(Z)$, and $\lambda=E(k)$. These have to be estimated and they determine the variances $V(\varepsilon_{a,1})=V(\varepsilon_{a,2})$ and $V(\xi)$.

A potential danger is that the model would not be identifiable, which means that different parameterisations of the random mechanism could result in the same distribution of difference $\xi$ of Eq. (2). For example, a normally distributed random variable can be the sum of extremely different random variables, and there are an infinite number of opportunities to formulate such a sum. However, in the current instance, the mechanism is identifiable because the empirical distribution of the sample of the following section has a special, and not a normal shape.

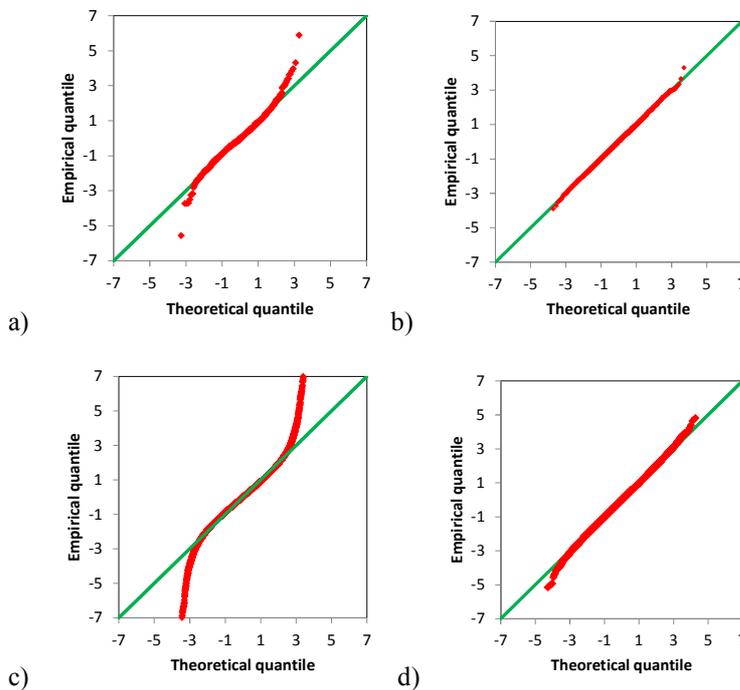

Fig. 3: Q-Q normal plots: a) for the difference $\xi$ of the observations (see section 3), b) of sample ($n$=10,000) of $Y_1$ and $Y_2$ of the autoregressive process, generated by a Monte Carlo simulation (200 realisations in every simulated sequence of the autoregressive process), c) from sample ($n$=100,000) of random impulses with $\lambda$=7.9, generated by a Monte Carlo simulation, d) the same as c) with $\lambda$=16.

The Q-Q normal plot (Upton and Cook, 2008) of this sample is shown in Fig. 3a for a normal distribution. If the plot is approximately a straight line, the random variable is (approximately) normally distributed. The observations of $\xi$ are obviously not a normal distribution, although it is similar to a normal distribution in a certain range. However, the upper and lower tails are completely different. A large sample of components was generated by a Monte Carlo simulation of the time history of the autoregressive process of Fig. 2b. The corresponding difference $\xi$ is (approximately or exactly) normally distributed according to the Q-Q normal plot of Fig. 3b. The Q-Q normal plot of Fig. 3c indicates the difference $\xi$ of components $Y_1$ and $Y_2$ that are generated with a Gumbel distributed $Z$ and a Poisson distributed $k$, with $\lambda$=7.9; the difference $\xi$ is not normally distributed and the Q-Q normal plot is similar to that of the real data of Fig. 3a. The difference $\xi$ would be more normally distributed, if the intensity $\lambda$=16 (Fig. 3d). Clearly, the random mechanism of Eq. (3) can approximate the generation of components $\varepsilon_{a,1}$ and $\varepsilon_{a,2}$.

## 3   The analysed sample

I used the PGA data of the SHARE project (Giardini, 2013; Share_Metafile_v3.2a.xls) for the current research on $\varepsilon_a$. Only observations were considered that included both horizontal components, were explicitly free field observations (column of Share_Metafile_v3.2a.xls: STRUCTURE TYPE), with information about the moment magnitude. The station locations are indicated in Fig. 4a, while the epicentres are shown in Fig. 4b. The corresponding sample of the random difference $\xi$ has the size $n$=1,829, and is shown in Fig. 3a. The difference of the sample mean to zero is small and insignificant. The estimated variance is $\hat{V}(\xi)$=0.1193.

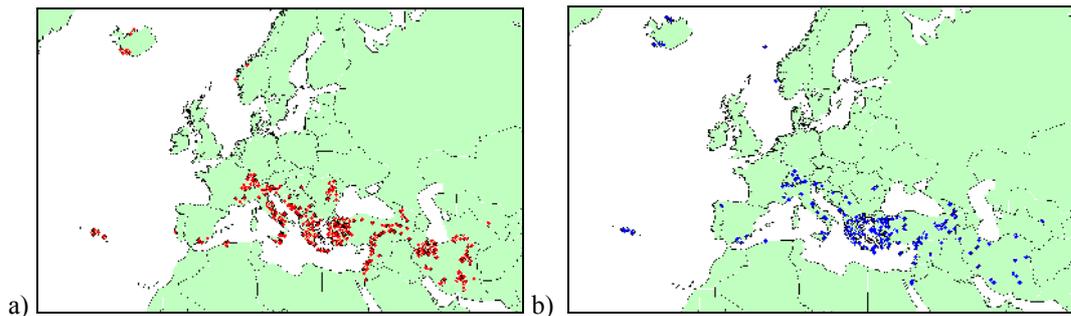

Fig. 4: Geographic information about the analysed sample of two horizontal components of PGA and the corresponding difference $\xi$, the samples are of regions in Europe and the Middle East: a) Map of the 707 recording stations, b) Epicentres of the 498 recorded earthquakes

## 4   Model building and estimation procedure

### 4.1   Basic parameter estimation

In the estimation, the parameters $E(Z)$ and $V(Z)$ were adapted for a fixed Poisson intensity $\lambda$, ensuring that $E(\varepsilon_{a,1})= E(\varepsilon_{a,2})$=1 and $V_{sample}(\xi)=V_{model}(\xi)$. This is done by a simple numerical optimisation, in which the value $((E(\varepsilon_a)-1)^2+(V_{model}(\xi)/V_{sample}(\xi)-1)^2)$ is minimised. The condition $V_{sample}(\xi)=V_{model}(\xi)=\hat{V}(\xi)=0.1193$ implies a moment estimator (also called method of moments [see, e.g., Soong, 1969]). $E(\xi)=0$ applies for all parametrisations. The remaining parameter is the Poisson intensity $\lambda$, which is estimated by the well-known maximum likelihood (ML) method (see, e.g., Lindsey 1996). The likelihood function $L(\theta)$ is formulated with the probability density function (PDF) $f_\xi$ of the random difference

$$L(\boldsymbol{\theta}) = \prod_{i=1}^{n} f_\xi(\xi_i; \boldsymbol{\theta}) \qquad (4)$$

where $\boldsymbol{\theta}$ is the parameter vector that only includes $\lambda$ in this instance, $n$ is the sample size, and $\xi_i$ is an element of the sample. $L(\boldsymbol{\theta})$ has to be maximised in the ML estimation.

Unfortunately, the formulation for PDF $f_\xi$ is not known. However, for the fixed parameters $\lambda$, $E(Z)$, and $V(Z)$, an extremely large sample of $\xi$ can be generated by a Monte Carlo simulation. The corresponding PDF can be approximated by the Kernel density estimation (also called Kernel smoothing), according to Silverman (1998). In this density estimation, a Kernel function $K$ is applied and the PDF is approximated with

$$f_\xi(x) \approx \sum_{i=1}^{m} \frac{K((x-x_i)/h)}{mh} \qquad (5)$$

where $h$ is the bandwidth and $x_i$ is the realisation of the simulated sample of size $m$. The standard normal distribution (cf. appendix) was applied as the Kernel function. The bandwidth was defined with $h=1.06 V_{observed}(\xi)^{0.5} m^{-0.2}$, as it is the optimal bandwidth for a normal distribution (Silverman, 1998), and the distribution of $\xi$ was similar to a normal distribution in a certain range (cf. Fig. 3a). Keef et al. (2009) have already used such an approximation of a distribution by a Monte Carlo simulation for the statistical modelling of river floods. Furthermore, Kernel smoothing is already being used in seismology, e.g., by Kijko (2004).

### 4.2 Estimation for the instance of independent variances

The sample size $m=100,000$ was chosen to approximate $f_\xi$ by the Monte Carlo simulation. Furthermore, the likelihood was computed for different $\lambda$, with increment 0.1. The random generator starts with the same starting value for every parametrisation to ensure the stability of the estimation. That is why a random impulse and angle were generated 200 times for every simulated realisation of $\varepsilon_{a,1}$ and $\varepsilon_{a,2}$, although only $k$ impulses and angles were applied.

The likelihood function of Eq. (4) has a maximum at $\hat{\lambda}=7.9$, and the corresponding parameter estimations are listed in Table 1 (column Gumbel distribution). It must be pointed out that the conditions $E(\varepsilon_{a,1})=E(\varepsilon_{a,1})=1$ could not be fulfilled for extremely small values of $\lambda$<5, because of the limit $V(Z) \geq 0$. Furthermore, the approximation of the PDF $f_\xi$ by a Monte Carlo simulated sample includes a certain inaccuracy, which also results in the local maxima of the likelihood function.

I outline that the variances $V(\varepsilon_a)$ and $V(\xi)$ do not depend on any further variable (e.g. magnitude) in this procedure.

### 4.3 Alternatives without dependence

The log-normal and the gamma distribution (cf. appendix) were considered as alternative distributions for $Z$. The corresponding estimations are $\hat{\lambda}$=7.9 and 8.1. The corresponding parameters are listed in Table1 and the likelihood functions are shown in Fig. 5a. The estimations of $V(\varepsilon_a)$ differ less, while the estimations of $\sigma_a$ are almost equal. The Gumbel distribution results in the largest $V(\varepsilon_a)$, the only factor I considered in further research.

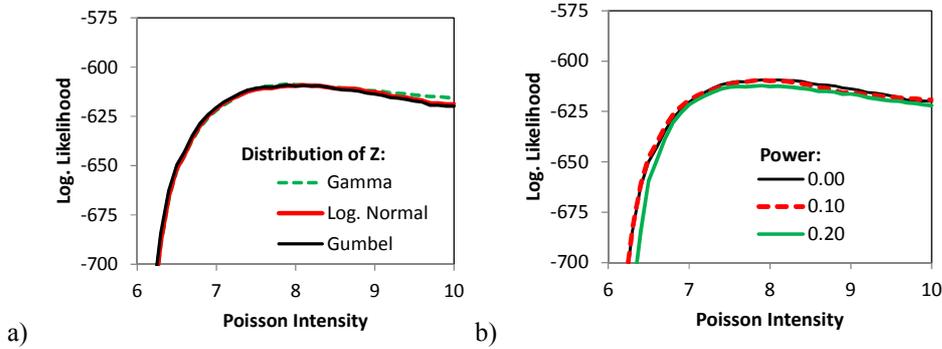

Fig. 5: Likelihood functions: a) results for different distributions for impulse $Z$, b) results for different alternatives according to Eq. (6)

Table1: Estimated parameters for different distribution models for random impulse $Z$

| Distribution of $Z$ | Estimated parameters | | | Corresponding parameters | |
|---|---|---|---|---|---|
| | $\lambda$ | $E(Z)$ | $V(Z)$ | $V(\varepsilon_a)$ | $\sigma_a$ of $\ln(\varepsilon_a)$ |
| Gumbel | 7.9 | 0.879 | 0.0497 | 0.0575 | 0.2518 |
| Log-normal | 8.1 | 0.862 | 0.0558 | 0.0565 | 0.2517 |
| Gamma | 7.9 | 0.866 | 0.0577 | 0.0522 | 0.2510 |

Additionally, an alternative was considered for Eq. (3a), with the power parameter $\alpha$ in

$$\varepsilon_a(w) = \max(Z_1|\cos(w-v_1)|/1^\alpha, Z_2|\cos(w-v_2)|/2^\alpha, ..., Z_k|\cos(w-v_k)|/k^\alpha), \alpha \geq 0 \quad (6)$$

The corresponding likelihood functions are presented in Fig. 5b. There was no significant improvement, since, according to the Bayesian information criterion of Schwarz (1978), the preference was for the original variant, without the additional parameter $\alpha$.

### 4.4 Distribution of the random component $\varepsilon_a$ and model validation

The empirical distributions (cf. appendix) of the observed and simulated sample of the difference $\xi$ were compared to validate the modelling. As indicated by Figs 6a and b, the distributions were quite similar, and the lower tail had the same shape as the upper tail because the distribution was symmetric. No quantitative test was conducted because the excellent visual match. There also was no special goodness-of-fit test for the specific situation (cf. Stephens 1986).

The simulated distribution of the random component $\varepsilon_a$ was also researched; however, a distribution model that fitted well was not found. The log-normal distribution and the Gumbel distribution (cf. appendix) poorly approximated the upper tail (Fig. 6c). The lower tails differed completely (Fig. 6d). Nevertheless, the approximations were considered in section 5.

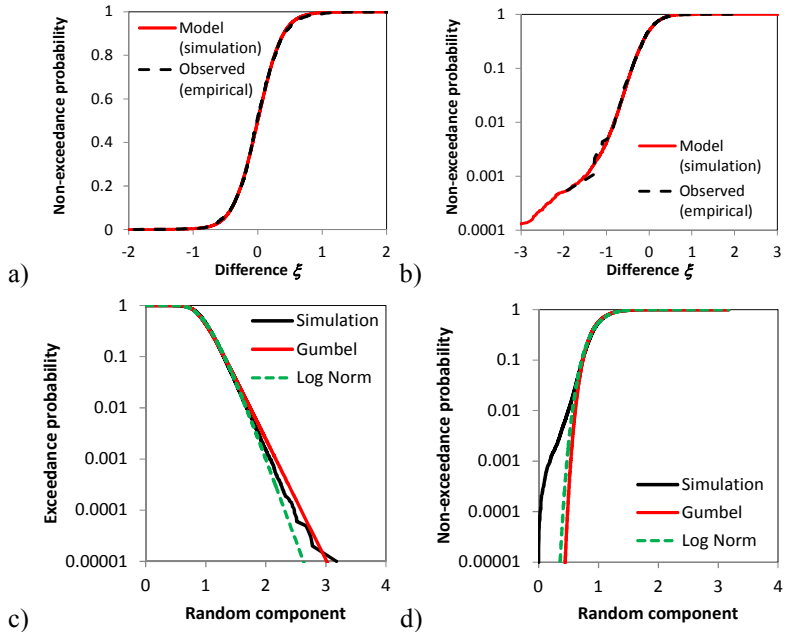

Fig. 6: Comparison of distributions: a) distribution of random difference $\xi$, b) distribution of random difference $\xi$, but with a logarithmised scale, c) upper tail of distribution models for $\varepsilon_a$, d) lower tail of distribution models for $\varepsilon_a$.

### *4.5 Analysis of potential dependencies*

The dependence of variance $V(\xi)$ on the magnitude and distance variables was analysed. For this purpose, the relation $V(\xi)=E(\xi^2)$ was considered, because $E(\xi)=0$ according to Eq. (2), while the additive regression to $ln(\xi^2)$ was applied as the predicted variable. In linear additive regression analysis, the best prediction variable is selected first and the corresponding regression model is subsequently estimated. The corresponding residuals are the predicted variables in the next step, and the best predictor is selected from the remaining potential predictors. The corresponding regression model is estimated once again, and the process is repeated. The details are explained by Fahrmeier et al. (2013, section 13). This procedure circumvents the problem of collinearity.

The Pearson correlation coefficient was applied as the selection criterion for the linear models. The logarithmised distances are used as predictors to ensure linearity. The instance $ln(0)$ for the epicentre distance $D_E$ and the Joyner-Boore distance $D_{JB}$ were eliminated by a simple modification $D_E^* = \sqrt{(D_E^2+1)}$ and $D_{JB}^* = \sqrt{(D_{JB}^2+1)}$. In instances where there was no $D_{JB}$ in a record, $D_{JB} = D_E$ was used. The correlation coefficients for the different distances are listed in Table 2. Some distance variables include the random hypocentre depth $H$. The epicentre distance $D_E^*$ is the best predictor for the first step of additive regression. The corresponding regression model is presented in Eq. (7). The resulting residuals have no significant correlation to the remaining potential predictors, as indicated by Table 2. The results presented in Table 3 confirm that the magnitude does not have any significant influence on the variance $V(\xi)$.

$$E(\ln(\xi^2)|D_E^*) = \alpha^* \ln(D_E^*) + \beta^*, \; \alpha^* = -0.2518, \; \beta^* = -2.6199 \tag{7}$$

Table 2: Estimated correlation coefficient $r$ (bold text: best correlation with highest value $r^2$; italic text: no statistical significance because $r=0$ is in the corresponding 90% confidence interval, according to Johnson [1995, p.576])

| Predicted | Potential predictors | | | | |
|---|---|---|---|---|---|
| | $\ln(D_E^*)$ | $\ln(\sqrt{D_E^2+H^2})$ | $\ln(D_{JB}^*)$ | $\ln(\sqrt{D_{JB}^2+H^2})$ | $M_w$ |
| $\ln(\xi^2)$ | **-0.1241** | -0.1158 | -0.1235 | -0.1147 | -0.0889 |
| Residuals of Eq. (7) | - | *0.0051* | *-0.0020* | *-0.0049* | ***-0.0311*** |

Table 3: Estimations of variance $V(\xi)$ for different ranges of distances and magnitudes (number of observations in brackets)

| # | >Min $D_E^*$ | ≤Min $D_E^*$ | $V(\xi)$ with $M_w \leq 5.7$ | $V(\xi)$ with $M_w > 5.7$ |
|---|---|---|---|---|
| 1 | 0 | 35 | 0.1518 (564) | 0.1513 (141) |
| 2 | 35 | 65 | 0.1449 (231) | 0.1012 (130) |
| 3 | 65 | 127 | 0.0621 (174) | 0.0773 (188) |
| 4 | 127 | without limit | 0.0890 (156) | 0.0903 (207) |

I have used the $\ln(\xi^2)$ and $\ln(D_E^*)$ because the relation is approximately linear; however, as the prediction of $Var(\xi)$ was needed, the non-linear regression model was estimated

$$E(\xi^2|D_E^*) = \beta D_E^{*\alpha}, \; \alpha = -0.2401, \beta = 0.2862, V(\xi|D_E^*) = E(\xi^2|D_E^*) \tag{8}$$

The least squares estimator was applied in a numerical optimisation to estimate the parameters. Based on the transformation between Eq. (7, 8), the power $\alpha$ was close to the slope $\alpha^*$. Factor $\beta$ is determined by $\beta = E(\varepsilon^2)/E(D_E^{*\alpha})$ (refer to Fig. 7a for the details).

Eq. (8) was implemented in the impulse model by a larger computation. The relation between $V(\xi)$ and $E(Z)$, $V(Z)$ and $V(\varepsilon_a)$ was computed for the Poisson intensity $\lambda=7.9$, under the aforementioned conditions of $E(\xi)=0$ and $E(\varepsilon)=1$. The resulting relations are depicted in Figs 7b and c. The parameters of $Z$ and $\varepsilon$ were determined indirectly in relation to the distance. Subsequently, the corresponding distribution of $\xi$ was generated with Monte Carlo simulations (Fig. 7e), taking into account the distance distribution of the data example (Fig. 7d). The model with distance dependence also performed exceptionally well.

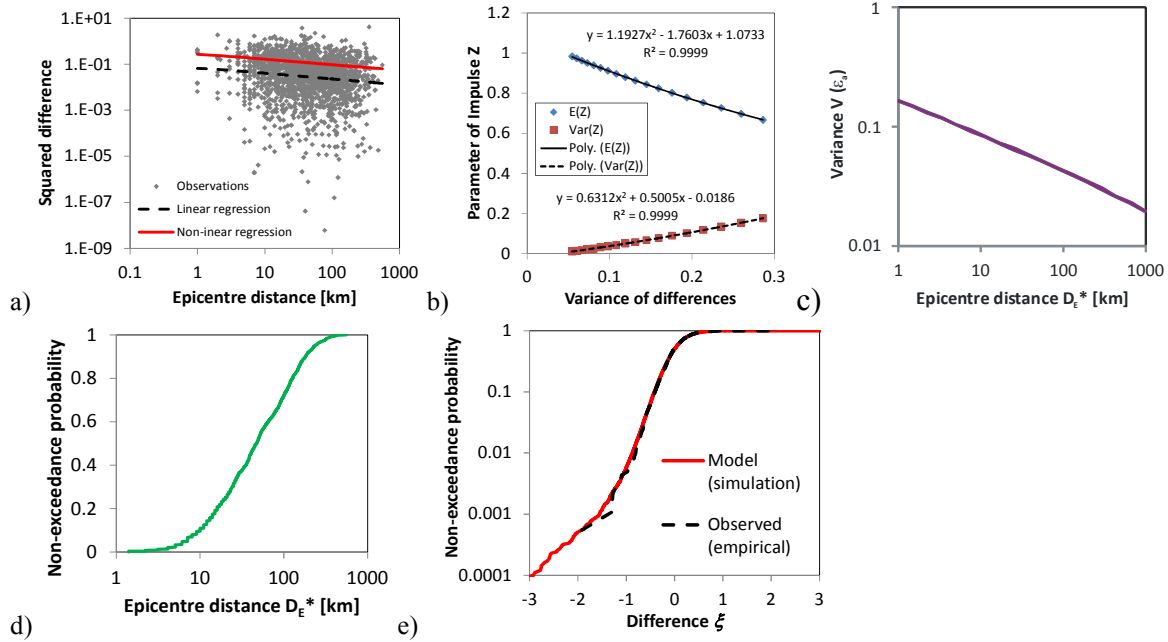

Fig. 7: Details of distance dependence: a) relation between epicentre distance $D_E^*$ and $\xi^2$, b) parameter of the $Z$ in dependence of $V(\xi)$, c) corresponding relation between distance and $V(\varepsilon_a)$, d) empirical distribution of the modified epicentre distance, e) validation of the model, with distance dependence by the distributions of $\xi$.

## 5 Influence on the PSHA

The result of PSHA could be influenced quite considerably by the random component $\varepsilon_a$ and its parameters, which is why the influence of the new models on the results of PSHA was researched. Additionally, I wanted to check whether the approximation of the distribution of $\varepsilon_a$, according to section 4.4, would succeed. For this purpose, a simple situation was constructed, featuring a source region with homogenous seismicity in time and space. The relation was estimated between the ground-motion level and the average of the annual exceedance frequency for the location in the centre of this source region (Fig. 8a). The assumed annual exceedance frequency of the magnitudes of the source region is shown in Fig. 8b. Only magnitudes $M \geq 4$ were taken into consideration. The upper bound magnitude was $m_{max}=6.5$. Furthermore, I set the hypocentre depth $H=20$ km in the simple GMR for $Y$ [PGA], which used the hypocentre distance $R$ [km] (Fig. 8c)

$$E(Y) = \exp(-2.4 + M_w - \ln(R) - 0.0005R), \quad R = \sqrt{D^2 + H^2}. \tag{9}$$

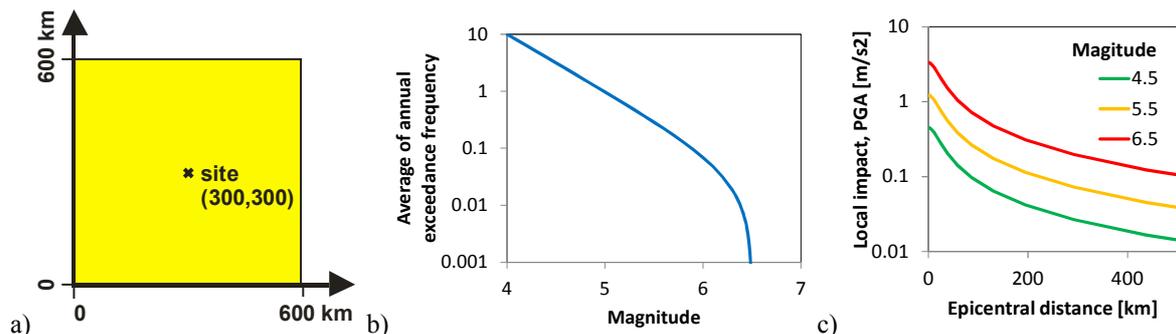

Fig. 8: Example of PSHA: a) analysed site and considered source region, b) seismicity of the source region ($M$ follows a truncated exponential distribution, cf. Cosentino et al., 1977), c) assumed GMR

The inter-event variability $\varepsilon_b$ is a log-normal distribution and has a variance of $V(\varepsilon_b)=0.4190$ that implies $\sigma_b=0.3606$. The latter is in the typical dimension of conventional estimations (cf. Abrahamson and Silva, 2008, Table 6, 1st row, columns $s_3$ and $s_4$).

In contrast with the conventional procedures (cf. Cornell, 1968; McGuire, 1995), I used a Monte Carlo simulation to compute the PSHA. Assatourians and Atkinson (2003) have already suggested Monte Carlo simulations for PSHA. Random events for one million years were generated for the research. The advantage of the Monte Carlo simulation is that the individual random component $\varepsilon_a$ can be simulated exactly according to the stochastic mechanism. In addition, approximations were considered, a Gumbel distribution was applied, as well as log-normal distributions for simulation $\varepsilon_a$. These distributions were parameterised by expectation $E(\varepsilon_a)=1$ and variance $V(\varepsilon_a)$, which were estimated in section 4. Consequently, three variants for generating $\varepsilon_a$ were considered, while, for each of these, two variants were considered, namely, with or without distance dependence.

The resulting averages of the annual exceedance frequencies for a defined PGA are shown in Fig. 9. The approximation of the distribution for $\varepsilon_a$ by a Gumbel or log-normal distribution functioned well for my example and up to a return period of 10,000 years. The difference found for the larger return periods could be based on the inaccuracies of the Monte Carlo simulation. The parameter dependent on the epicentre distance resulted in a higher exceedance frequency for a defined PGA, compared with the variants with constant parameters and $\sigma_a \approx 0.3$ (visual estimation, according to the result for a return period of 10,000 years). The contribution of the inter-event variability $\varepsilon_b$ to the hazard was larger than the contribution of $\varepsilon_a$, which is in contrast to the previous parameterisations with $\sigma_a > \sigma_b$ (see, e.g., Abrahamson and Silva, 2008, Table 6). The conventional modelling leads to considerable larger hazard for large return periods (compare Fig.9a and b with c).

It is emphasised that the influence of the random component strongly depended on the upper bound magnitude, and it was equal to the instance of $\sigma_a=0$ if an infinite upper bound of exponentially distributed magnitudes were present. The latter is a result of extreme value statistics (cf. Schlather, 2002, theorems 1 and 2), as was mentioned already for PSHA by Raschke (2013a).

The truncation of the combination random components $\varepsilon_a\varepsilon_b$ and $ln(\varepsilon_a)+ln(\varepsilon_b)$ was not considered, in contrast with previous PSHA (e.g. Bommer et al., 2004; Strasser et al., 2008). There are different concerns in truncated random components, such as the statistical estimation or physical deviation of the truncation point. In addition, the truncation of the combination needs the truncation of every single random component. To my knowledge, this factor has not been explained or discussed before.

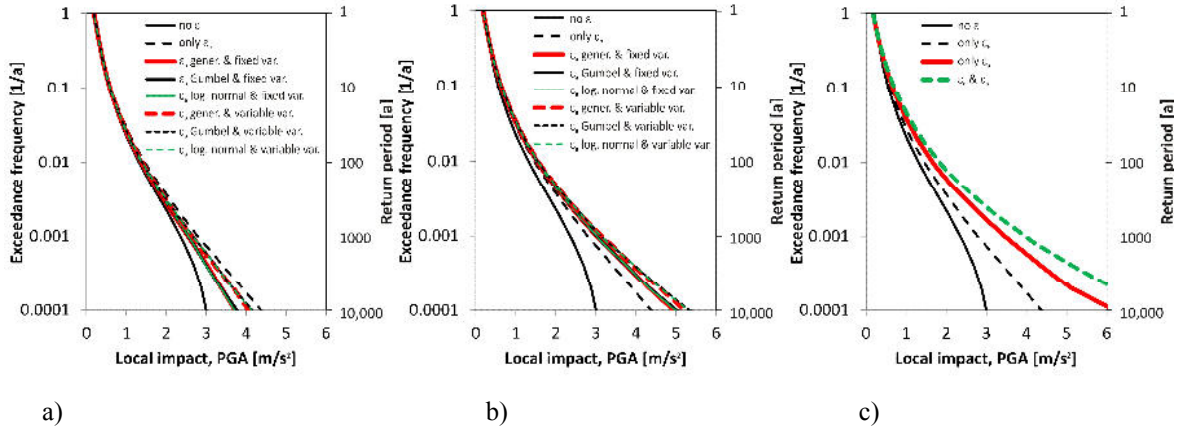

a)                          b)                          c)

Fig. 9: Estimated hazard curves: a) without the inter-event variability, b) with inter-event variability, c) $\varepsilon_a$ according to untruncated conventional models with $\sigma_a$=0.55 (for comparison, instances without any $\varepsilon$ and with only $\varepsilon_b$ are shown in both figures)

## 6 Conclusion, discussion and outlook

I have formulated a simple stochastic mechanism by employing Eq. (3). This mechanism is a sequence of random impulses, with a random direction in the plane of the horizontal components that generates the individual random component $\varepsilon_a$ of Eqs (1–2). This approach is the first one, which explicitly and satisfactorily reproduces the observed distribution of the difference $\xi$ of the logarithmised horizontal components relevant to the local ground-motion intensity (Figs 6 and 7). The corresponding estimation is $\sigma_a$=0.2518 (standard deviation of $ln(\varepsilon_a)$, without dependence on the epicentre distance, cf. Table 1) and it is not sensitive to the distribution assumption for the random impulse. The variant with dependence on the epicentre distance results in a hazard curve for the example in section 5, which is approximately equivalent to an independent model with $\sigma_a$≈0.30. Both values of $\sigma_a$ are small compared with the values of conventional GMRs. For example, Abrahamson and Silva (2008) published estimations in the range 0.46 to 0.60. However, they have not considered area-equivalence or event-specific GMRs in their estimation for the individual random component. The average variance of the event-specific GMRs of Raschke (2013a) corresponded to $\sigma_a$=0.436 and the smallest variance with $\sigma_a$=0.33, although regression analysis was applied. This indicates that the dimension of the parameter estimation was reasonable. Furthermore, the resulting distribution of the individual random component can be approximated in a certain range by a Gumbel or log-normal distribution.

The results of this novel approach are promising in light of the importance of the individual random components in PSHA (cf. Bommer et al., 2004). Therefore, I recommend further validation by additional research, such as the application of the model and methods for the samples of different regions. Another important validation would be the examination of the distribution of other corresponding random variables, such as the quotient of a horizontal component and the absolute maximum over all directions in the horizontal plane. In addition, the ratio between the area and the squared perimeter of the geometric figure drawn by *Y(w)* of Eq. (3b) (c.f. Fig. 2, red lines) could be a relevant random variable. In all instances, the influence of all other elements of GMR according to Eq. (1) has to be eliminated. Only the individual random component should determine the researched random variables.

Furthermore, the dependence of variances $V(\xi)$ and $V(\varepsilon_a)$ on epicentre distance should be examined in future research. The physical interpretation of this effect should be formulated and justified. It must be pointed out that the decreasing of $V(\varepsilon_a)$ and $\sigma_a$ by increasing distances differs from some previous GMR, which included a decreasing $\sigma_a$ (c.f. Abrahamson et al., 2008, Fig. 11).

To conclude, the dependence between the horizontal components $Y_1$ and $Y_2$ and $\varepsilon_{a,1}$ and $\varepsilon_{a,2}$ should be researched, as it could be important for appropriately designing the seismic resistance of buildings or nuclear facilities.

## Acknowledgements

I thank my colleagues Andrzej Kijko and Vasily Pavlenko, who provided insight and expertise that significantly assisted us in my research and the writing of this paper.

# Appendix

*Distribution models and corresponding functions*

The probability density function (PDF) of the log-normal distribution of a real valued random variable X is (cf. Johnson et al., 1994, section 14)

$$f(x) = \frac{1}{\sqrt{2\pi}\sigma x} \exp\left(-\frac{(\ln(x)-\mu)^2}{2\sigma^2}\right), \quad X > 0 \tag{A1}$$

and includes the following moments

$$E(\ln(X)) = \mu \tag{A2}$$

$$V(\ln(X)) = \sigma^2 \tag{A3}$$

$$E(X) = \exp(\mu + \sigma^2/2) \tag{A4}$$

and

$$V(X) = \exp(2\mu + \sigma^2)(\exp(\sigma^2) - 1) \tag{A5}$$

The PDF of standard normal distribution is (cf. Johnson et al., 1994, section 13)

$$f(x) = \frac{1}{\sqrt{2\pi}} \exp\left(-\frac{x^2}{2}\right), \quad X > 0 \tag{A6}$$

The cumulative distribution function (CDF) of the log-normal and the normal distribution cannot be expressed by a simple equation, but has to be computed numerically (e.g., by a function of MS Excel).

The Gumbel distribution of a real valued random variable X has CDF (cf. Johnson et al., 1995, section 22)

$$F(x) = \exp(\exp(-(x-a)/b)) \tag{A8}$$

and has the moments

$$E(X) = a + b\gamma, \quad \gamma \approx 0.57722 \tag{A9}$$

and

$$V(X) = b^2 \pi^2 / 6 \tag{A10}$$

and

The gamma distribution of a real valued random variable X has the PDF (cf. Johnson et al., 1994, section 17)

$$f(x) = b^a x^{a-1} \exp(-bx)/\Gamma(a) \tag{A11}$$

where $\Gamma$ is the elementary gamma function. The moments are

$$E(X) = a/b \tag{A12}$$

and

$$V(X) = a/b^2 \tag{A13}$$

The Poisson distribution for a discrete random variable k≥0 is formulated by (cf. Johnson et al., 2005, section 4)

$$P(k=x) = \frac{\lambda^x \exp(-\lambda)}{x!} \tag{A14}$$

and has the moments

$$E(k) = V(k) = \lambda \tag{A14}$$

The empirical distribution function of a random variable *X*, with sorted sample *($X_1 \leq X_2 \leq ... \leq X_i \leq ... \leq X_n$)* is

$$\hat{F}(X_i) = i/(n+1) \tag{A15}$$

because the expectation is *E(F($X_i$))=i/(n+1)* (see, e.g., David and Nagaraja, 2003; Ahsanullah et al., 2013, chapter 2).